# On the Kurtosis of Modulation Formats for Characterizing the Nonlinear Fiber Propagation

Junho Cho, *Senior Member, IEEE*, and Robert Tkach

*Abstract*—Knowing only two high-order statistical moments of modulation symbols, often represented by the fourth moment called "kurtosis", the overestimation of nonlinear interference (NLI) in a Gaussian noise (GN) model due to Gaussian signaling assumption can be corrected through an enhanced GN (EGN) model. However, in some modern optical communication systems where the transmitted modulation symbols are statistically correlated, such as in systems that use probabilistic constellation shaping (PCS) with finite-length sphere shaping, the kurtosis-based EGN model produces significant inaccuracies in analytical prediction of NLI. In this paper, we show that for correlated modulation symbols, the NLI can be more accurately estimated by substituting a statistical measure called windowed kurtosis into the EGN model, instead of the conventional kurtosis. Remarkably, the optimal window length for windowed kurtosis is found to be consistent with the self-phase modulation (SPM) and cross-phase modulation (XPM) characteristic times in various system configurations. The findings can be used in practice to analytically evaluate and design NLI-tolerant modulation formats.

*Index Terms*— Nonlinear fiber propagation, kurtosis, probabilistic constellation shaping, enhanced Gaussian noise model.

## I. INTRODUCTION

NONLINEAR interference (NLI) can be accurately modeled as Gaussian noise (GN) in long-haul dispersion-uncompensated fiber links [1]–[6]. Treating the NLI as GN allows (semi-) analytical estimation of the signal-to-noise ratio (SNR) of recovered signals after nonlinear fiber propagation, without resorting to time-consuming split-step simulations [7]–[8]. The GN model is accurate when the modulation symbols are continuous Gaussian. However, all practical systems use discrete alphabets, such as quadrature amplitude modulation (QAM) alphabets with or without probabilistic constellation shaping (PCS), not a continuous alphabet. In such systems, the NLI can still be quantified with high accuracy by using an enhanced GN (EGN) model [4]–[6] which introduces correction terms in the GN model to remove the overestimation of NLI due to the Gaussian signaling assumption. This correction can be accomplished simply by substituting into the EGN model the 4th and 6th statistical moments of the underlying modulation format. The high-order statistical moments are often represented by the 4th moment called *kurtosis* in many papers since the 4th moment has the greatest influence on the modulation-dependent NLI among all high-order moments.

For this reason, designing modulation formats to have small kurtosis has been an important approach to reduce NLI [9]–[12]. In [9], a heuristic short-length sphere shaping method is used that yields a smaller kurtosis than long sphere shaping at the expense of increased shaping rate loss. In [10], the kurtosis of a modulation format is reduced by allowing only constant-modulus symbols created in a 4-dimensional (4D) signal space, i.e., by making all polarization-division multiplexed (PDM) symbols defined in the modulation format have the same power in each time slot. In [11], the kurtosis is reduced by removing the symbol sequences that have the largest deviation from the average power among all possible symbol sequences. In [12], a low-kurtosis modulation format is constructed by imposing a constraint on the maximum kurtosis of symbols in a sequence, in addition to the constraint on the maximum variance (i.e., power, or energy) of the symbols in the sequence as in conventional sphere shaping [13]–[18].

However, the kurtosis-based modulation format design is *not* fully supported by theoretical foundation in systems where the modulation symbols are statistically correlated, such as in systems that use PCS [19] with finite-length sphere shaping [9], [11], [12], [14]–[18], or in systems utilizing multi-dimensional modulation formats [10], [13], [20]–[22]. This is because in such systems, the assumption of the EGN model that the modulation symbols are statistically independent of each other does not hold, which is essential to arrive at the conclusion that the NLI increases with kurtosis. There are several recent papers [16], [22]–[24] showing empirical evidence that even at the same kurtosis, NLI changes when the temporal correlation between modulation symbols changes. However, for the systems with correlated modulation symbols, there are no analytical nonlinear fiber propagation models known to date, except within four dimensions [25], [26].

As a clue to understanding the relationship between kurtosis and NLI for a multitude of correlated modulation symbols, it was first reported in [27] that there exists a high correlation between NLI and the energy dispersion of modulation symbols, when the energy is measured using a measurement *window* longer than a modulation symbol period. In [28], we showed that the NLI for statistically dependent modulation symbols can also be estimated analytically with higher accuracy than conventional methods, using the kurtosis measured with a measurement window longer than a modulation symbol period, which we call *windowed kurtosis*. We allow for mismatch between the EGN model and the true system caused by the statistical dependency of modulation symbols, but compensate for inaccuracies in NLI prediction by substituting the windowed kurtosis into the EGN model instead of the conventional kurtosis. Remarkably, the measurement window length for the windowed kurtosis is given as a function of the symbol rate, signal bandwidth, and transmission distance, and this coincides with the self-phase modulation (SPM) and cross-phase modulation (XPM) characteristic times.

In this work, we show that our previous findings [28] apply to more general systems. The measurement window length for windowed kurtosis obtained in wavelength-division multiplexing (WDM) systems with a total bandwidth of 100 GHz [28] also applies to WDM systems with a wider total bandwidth of 500 GHz used in this paper, which suggests that the measurement window length of [28] may be generally applicable to WDM systems of any total bandwidth. We further show that the windowed kurtosis leads to better EGN model predictions for multi-dimensional symbols as well as one-dimensional symbols, suggesting that the windowed kurtosis can be used to characterize a wider class of correlated modulation.

## II. Kurtosis

Without loss of generality, we can assume that complex-valued transmit signal has zero mean since otherwise the modulation alphabet can be shifted by the mean value (i.e., so called the direct current (DC) term) to minimize the average signal power. Since we deal with only PDM systems in this work, we consider that the transmit signal $x$ is 4D. When evaluating the nonlinear fiber propagation effects, it is convenient to normalize $x$ to have a unit power as

$$u = \frac{x}{\sqrt{\langle \|x\|^2 \rangle}}, \qquad (1)$$

where $\langle \cdot \rangle$ denotes statistical averaging and $\|\cdot\|$ denotes Euclidean norm. Then, we have that $\langle u \rangle = 0$ and $\langle \|u\|^2 \rangle = 1$ by assumption. The $k$-th *standardized moments* of the original signal $x$ are obtained from the normalized signal $u$ as

$$\mu_k \triangleq \langle \|u\|^k \rangle. \qquad (2)$$

The 4th moment $\mu_4$ is commonly referred to as *kurtosis*. Of all high-order moments, the kurtosis has the greatest influence on NLI, as discussed below.

The EGN model states that the average NLI power increases in proportion to the cube of the average signal power as

$$\langle P_{NLI} \rangle = \eta \langle \|x\|^2 \rangle^3, \qquad (3)$$

where $\eta$ denotes the NLI coefficient that is mostly determined by the link parameters such as the dispersion coefficient, channel bandwidth, span length, and the number of spans, and to some extent by the modulation format. According to [4], the NLI coefficient for SPM can be calculated as

$$\eta_{SPM} = \kappa_1 + \underbrace{(\mu_4 - 2)\kappa_2 + (\mu_6 - 9\mu_4 + 12)\kappa_3}_{(A)}, \qquad (4)$$

where $\kappa_1, \kappa_2, \kappa_3$ are functions of the link parameters and spectral shape of the transmit signal $x$. The first term $\kappa_1$ on the right-hand side quantifies the NLI coefficient for continuous Gaussian signals, and the term $(A)$ is a negative-valued correction term (hence reducing the NLI estimate) for non-Gaussian signals. In the term $(A)$, $\kappa_2$ is greater than $\kappa_3$ by an order or two in standard single mode fiber (SSMF) links, which explains the significance of $\mu_4$ over $\mu_6$ in determining the NLI. A mathematical expression of the NLI coefficient $\eta_{XPM}$ for XPM is not provided in this paper, since it consists of many more terms of $\kappa_1, \kappa_2, \kappa_3$ and $\mu_4, \mu_6$ than in (4), but it can be obtained in a similar way to (4) and is used for EGN model prediction in Section V; interested readers are referred to [4]–[6]. Note that throughout this paper, SPM refers to the NLI caused solely by spectral entities within the channel bandwidth (which corresponds to self-channel interference (SCI) of [4], see [4, eq. (5)]), and XPM refers to the NLI caused by simultaneous intervention of spectral entities within and outside the channel bandwidth (thus including both cross-channel interference (XCI) and multi-channel interference (MCI) of [4], see [4, eqs. (18) and (36)]).

In [28], we focused particularly on the deviation of the signal *power* from the mean value, rather than on the deviation of the signal itself from the origin as in (2). As will be discussed in Section IV and Appendix, this is necessary to find a new measure to replace kurtosis. To this end, let us define the normalized signal power as

$$\wp \triangleq \|u\|^2, \qquad (5)$$

which allows us to compactly describe the statistical moments. It immediately follows from (1) that $\langle \wp \rangle = 1$. The $k$-th *central moments* of $\wp$ are defined as the $k$-th order deviation of $\wp$ from its average; i.e.,

$$\mathfrak{m}_k \triangleq \langle (\wp - \langle \wp \rangle)^k \rangle = \langle (\wp - 1)^k \rangle. \qquad (6)$$

Then, the standardized moments of $x$ can be obtained solely from the central moments of $\wp$ as

$$\begin{aligned}\mu_4 &= \langle \wp^2 \rangle \\ &= \langle (\wp - 1)^2 \rangle + 2\langle \wp - 1 \rangle + 1 \\ &= \mathfrak{m}_2 + 2\mathfrak{m}_1 + 1,\end{aligned} \qquad (7)$$

and

$$\begin{aligned}\mu_6 &= \langle \wp^3 \rangle \\ &= \langle (\wp - 1)^3 \rangle + 3\langle (\wp - 1)^2 \rangle + 3\langle \wp - 1 \rangle + 1 \\ &= \mathfrak{m}_3 + 3\mathfrak{m}_2 + 3\mathfrak{m}_1 + 1.\end{aligned} \qquad (8)$$

In Section IV, an alternative statistical measure for the central moments (6) will be defined for correlated symbols, and the relations (7) and (8) will be used to substitute the new measures into the EGN model.

## III. Sphere Shaping

To understand why the conventional moments as defined in

(2) are not suitable for characterizing correlated symbols, we first need to know under what circumstances correlations arise between modulation symbols. Throughout this paper, we use sphere shaping [9], [11], [12], [14]–[18] for PCS, which is arguably one of the most important approaches to PCS from an information-theoretic perspective. In this section, we describe how sphere shaping creates correlations between symbols.

In traditional uniform PDM $M^2$-ary QAM systems *without* PCS, a transmitter (TX) produces modulation symbols with equal probabilities in each signal dimension (i.e., in each of the in-phase and quadrature components of a PDM symbol). Without loss of generality, the modulation symbols are selected from the alphabet $\mathcal{A}^\pm = \{\pm 1, \pm 3, ..., \pm(2M-1)\}$ in arbitrary units (a.u.). The 1D modulation symbols in this case can be seen as *independent and identically distributed (i.i.d.)* random variables with a uniform distribution in $\mathcal{A}^\pm$. The i.i.d. nature of the traditional QAM symbols is the key to enabling mathematical simplifications in analytic modeling of nonlinear fiber propagation by using the statistical moments as in (4).

### A. Sphere shaping over one-dimensional symbols

Sphere shaping yields a probability distribution of modulation symbols that is symmetric around the origin. This symmetry allows us to legitimately describe and analyze the sphere shaping with only the positive symbols drawn from the alphabet $\mathcal{A} = \{1, 3, ..., 2M-1\}$, so we omit the signs of symbols throughout this paper. Also, four *consecutive* sphere-shaped 1D symbols constitute one PDM symbol in this work, for reasons that will be explained in Section IV-A. For the convenience of description, we therefore assume that a shaping encoder produces $4n$-long 1D symbol blocks that are transmitted over $n$ time slots each.

Let us denote a block of sphere-shaped 1D symbols as $\boldsymbol{a} = [a_1, ..., a_{4n}]$ with $a_i \in \mathcal{A}$ for $i = 1, ..., 4n$, and the corresponding PDM symbol block as $\boldsymbol{x} = [x_1, ..., x_n]$ with $x_i = [a_{4i-3}, ..., a_{4i}] \in \mathcal{A}^4$ for $i = 1, ..., n$. When generating PDM symbol blocks, sphere shaping imposes a constraint on the total energy within a block such that

$$\|\boldsymbol{a}\|^2 = \sum_{i=1}^{4n} \|a_i\|^2 = \sum_{i=1}^{n} \|x_i\|^2 \leq E^*_{\text{Shaped}} \quad (9)$$

is fulfilled for a chosen energy limit $E^*_{\text{Shaped}}$. This method is called *sphere shaping* since if a shaped block is plotted as a $4n$-dimensional point, with its position on the $i$-th coordinate axis being $a_i$, the points are uniformly distributed over square lattice points on or within a uniform $4n$-dimensional (hyper-) sphere of radius $\sqrt{E^*_{\text{Shaped}}}$ (with the reflective symmetry due to equiprobable signs). The maximum amount of information bits that can be carried by a sphere-shaped symbol is given as

$$H = \frac{\lfloor \log_2 |\boldsymbol{a}| \rfloor}{n} \quad (10)$$

in bits per 1D symbol, where $|\boldsymbol{a}|$ denotes the number of all sequences $\boldsymbol{a}$ that fulfill the energy constraint (9).

Without sphere shaping, the maximum energy of a block is given as $E^*_{\text{Unshaped}} = \max_{\boldsymbol{a}} \|\boldsymbol{a}\|^2 = 4n(2M-1)^2$, which occurs when all $a_i$ in $\boldsymbol{a}$ have the maximum value $2M-1$. Sphere shaping uses $E^*_{\text{Shaped}} < E^*_{\text{Unshaped}}$, so among all blocks created in traditional QAM systems, the blocks with higher energies than $E^*_{\text{Shaped}}$ are prevented from being created by the shaping encoder. Fig. 1 shows the probability distribution of the square root of 1D signal power in three different 64-QAM systems. The figures on the left show the case where the signal power is measured per 1D symbol, and the figures on the right show the case where the signal power is measured per a block of twenty 1D symbols. When measured per 1D symbol, the square-root signal power in an i.i.d. uniform 64-QAM system (Fig. 1(a)) follows a uniform distribution, and the square-root signal power in a sphere-shaped 64-QAM system with shaping block length $n = 5$ (Fig. 1(b)) follows a Maxwell-Boltzmann (MB)-like distribution. As the shaping block length $n$ increases towards infinity (Fig. 1(c)), which represents an i.i.d. PCS 64-QAM system, the empirical distribution of the square-root signal power approaches an ideal MB distribution, and the shaping rate loss approaches zero, nearly achieving capacity in the additive white Gaussian noise (AWGN) channel. If the square-root signal power is measured per a block of twenty 1D symbols, the distributions become quite different. In cases where the modulation symbols are i.i.d. (Figs. 1(a) and 1(c)), the distribution tends toward a Gaussian distribution, but in the finite-length sphere shaping system (Fig. 1(b)), the distribution is concentrated near $\sqrt{E^*_{\text{Shaped}}/(4n)}$ due to the block-wise energy constraint (9) (this concentration can also be explained by a phenomenon called sphere hardening [29]); the blocks of symbols that have a greater power than this value is not created.

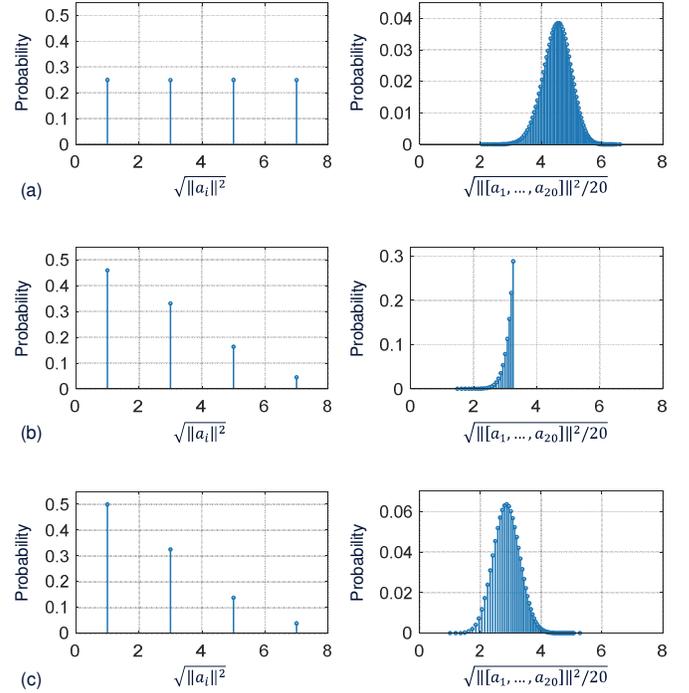

Fig. 1. Probability distribution of square root of 1D signal power in (a) i.i.d. uniform 64-QAM, (b) sphere-shaped 64-QAM ($n = 5$, $H = 1.6$), and (c) i.i.d. PCS 64-QAM ($H = 1.6$) systems. The figures on the left show when the signal power is measured in each 1D symbol, and the figures on the right show when the signal is measured in a block of twenty 1D symbols.

The correlations between sphere-shaped modulation symbols are in fact inherent in the sphere shaping principle since sphere shaping imposes a constraint that all symbols in a block must *jointly* satisfy. The correlation can also be illustrated by a simple example as follows. In a successive sphere shaping encoding process, suppose that the energy of previously generated symbols in a block has been accumulated close to $E^*_{\text{Shaped}}$. Then, high-energy symbols in $\mathcal{A}$ cannot be generated in the rest of the block. Therefore, the probability of occurrence of a symbol in a block changes depending on the symbols that have been generated before, or formally, $\Pr[a_i | a_1 \ldots a_{i-1}] \neq \Pr[a_i]$ for $i = 1, \ldots, 4n$. This shows that sphere-shaped symbols are *not* independent of each other but are correlated.

### B. Sphere shaping over multi-dimensional symbols

The shaping rate loss approaches zero as the block length $n$ increases, as mentioned above. However, as $n$ increases, the kurtosis of sphere-shaped 1D symbols also increases. So, if we apply the EGN model for i.i.d. symbols directly to correlated symbols, nonlinear optical fiber channels favor short block lengths for reduced kurtosis while linear channels favor long block lengths for reduced shaping rate loss, resulting in a fundamental tradeoff between the shaping rate loss and NLI. A possible approach to balance the shaping rate loss and kurtosis is to perform short sphere shaping to generate length-$N$ blocks of 1D symbols, with $N$ being a small number, and then treat each length-$N$ block as a single $N$-D symbol to perform long sphere shaping over the $N$-D symbols. The inner short shaping aims at reducing the kurtosis, and the outer long shaping attempts to reduce the shaping rate loss. In this work, we choose the inner short shaping length as $N = 4$ and create $x_i = [a_{4i-3}, \ldots, a_{4i}]$ for $i = 1, \ldots, n$ such that an inner shaping constraint

$$\|x_i\|^2 = \sum_{k=4i-3}^{4i} \|a_k\|^2 \leq E^*_{\text{Inner-shaped}} \tag{11}$$

is fulfilled for an energy limit $E^*_{\text{Inner-shaped}}$, in addition to the outer shaping constraint (9). This prevents occurrence of a large energy *in each time slot*. How much kurtosis is reduced thereby will be quantified in Section IV. For brevity, we refer to the conventional sphere shaping as 1D-shaping, and the inner-outer concatenated shaping described above as 4D-shaping. Note that due to the added constraint (11), 4D-shaping in some cases has to use a slightly larger $E^*_{\text{Shaped}}$ in (9) than 1D-shaping for the same $H$. As we will see in Section IV, this can put 4D-shaping at a disadvantage in terms of NLI. Among several implementations of sphere shaping, we use the enumerative sphere shaping (ESS) [15], [16] for both 1D- and 4D-shaping.

## IV. Windowed Kurtosis

Similarly to earlier studies on the finite-memory GN model [30] or the windowed measurement of the energy dispersion [27], [31], our previous study [28] begins with a question about the deficiency of the kurtosis in characterizing the correlated symbols. If the symbols are i.i.d. as in traditional QAM systems, the NLI does not change when the symbols are randomly permuted, nor does the kurtosis. On the other hand, if the symbols are correlated, the NLI *changes* under permutations of symbols across shaping blocks [23], but the kurtosis is still invariant. Therefore, in order to analytically determine the NLI for both the i.i.d. and correlated symbols, a new statistical measure is required, which can capture *local* properties of the signal in addition to the *asymptotical* properties of it. More specifically, the new statistical measure must fulfill the following conditions:

1) For i.i.d. symbols *without* local structure, the statistical measure must be the same whether obtained by global measurement or by averaging over local measurements.
2) For correlated symbols *with* local structures, the statistical measure must be different when measured with the above two methods. Furthermore, it must be able to quantify how substantial the local properties in the signal are.

### A. Measuring kurtosis with a window

In [28], we accounted for local properties of the signal by measuring the signal power for blocks of symbols, rather than for individual symbols. Namely, we used new statistical moments called *windowed central moments* of $\wp$, defined as

$$\overline{\mathfrak{m}}_k \triangleq \langle (\langle \wp \rangle_w - 1)^k \rangle \cdot \underbrace{(2w)^{k-1}}_{(B)} \tag{12}$$

for $k = 2, 3$, where $\langle \cdot \rangle_w$ denotes a moving average filter with window length $w$. While (6) quantifies the deviation of the signal power from the asymptotic average power by treating the power $\wp$ of each symbol as the signal power, (12) quantifies it by treating the power $\langle \wp \rangle_w$ measured over $w$ consecutive symbols as the signal power. We note that without the term (a) in (12), $\langle \wp \rangle_w$ approaches $\langle \wp \rangle = 1$ as $w$ increases, and thus the windowed central moments $\overline{\mathfrak{m}}_k$ approach 0 in the limit of $w \to \infty$ for all $n$. However, with the term $(B)$, it is ensured that $\overline{\mathfrak{m}}_2 = \mathfrak{m}_2$ and $\overline{\mathfrak{m}}_3 = \mathfrak{m}_3$ for all $w = 1, 2, \ldots$, *if* the symbols are i.i.d (see Appendix for proof). Namely, for i.i.d. symbols, the 2nd and 3rd windowed central moments do not change with the window length and are equal to the central moments measured without windowing, thus satisfying the condition for a good statistical measure for i.i.d. symbols. The factor 2 in (a) compensates for $\langle \wp \rangle_w$ being averaged over 2 polarizations. Note that without the term $(B)$, the 2nd windowed central moment $\overline{\mathfrak{m}}_2$ is the same as the energy dispersion index that was recently suggested in [27].

To obtain the NLI coefficient of a sphere-shaped modulation format using the EGN model as in (4), the windowed central moments $\overline{\mathfrak{m}}_k$ need to be converted back to the standardized moments $\mu_k$. This can be done by using the relations (7) and (8), but by substituting $\overline{\mathfrak{m}}_k$ in places of $\mathfrak{m}_k$; i.e., the *windowed standardized moments* of $x$ can be obtained from $\overline{\mathfrak{m}}_k$ as

$$\bar{\mu}_4 = \overline{\mathfrak{m}}_2 + 2\overline{\mathfrak{m}}_1 + 1, \tag{13}$$

and

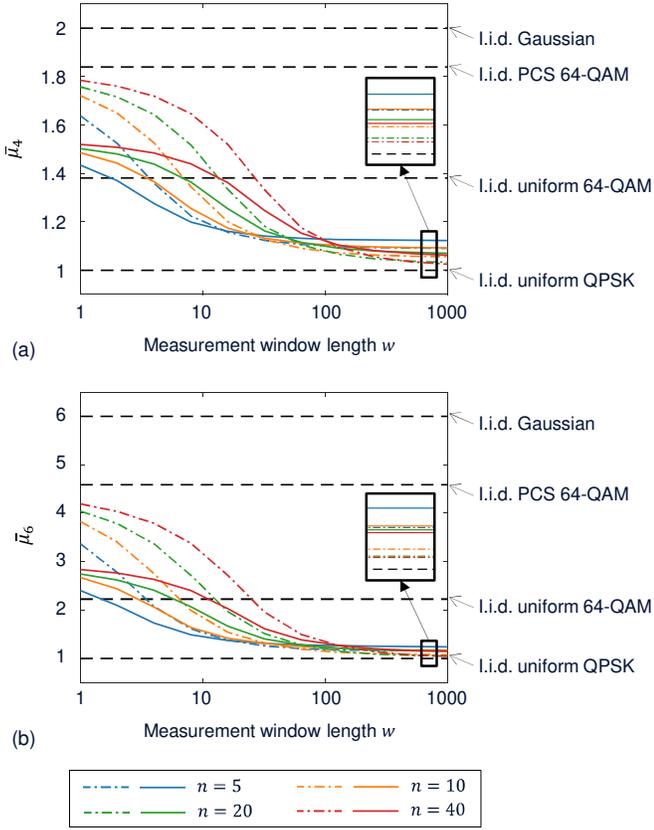

Fig. 2. (a) 4th and (b) 6th windowed standardized moments as a function of the measurement window length $w$. The dash-dotted lines represent 1D-shaped symbols, and the solid line represent 4D-shaped symbols.

$$\bar{\mu}_6 = \overline{m}_3 + 3\overline{m}_2 + 3\overline{m}_1 + 1. \qquad (14)$$

Correspondingly, the NLI coefficient for SPM is obtained from $\overline{m}_k$ instead of $m_k$ as

$$\eta_{SPM} = \kappa_1 + (\bar{\mu}_4 - 2)\kappa_2 + (\bar{\mu}_6 - 9\bar{\mu}_4 + 12)\kappa_3. \qquad (15)$$

Similarly, the NLI coefficient for XPM can be obtained from $\bar{\mu}_k$. Since $\overline{m}_2 = m_2$ and $\overline{m}_3 = m_3$ for i.i.d. symbols, we also have that $\bar{\mu}_4 = \mu_4$ and $\bar{\mu}_6 = \mu_6$ for i.i.d. symbols. Therefore, equations (13)-(15) represent general formulae for both i.i.d. and correlated symbols in nonlinear fiber propagation modeling, and equations (4), (7), and (8) can be seen as a special case where kurtosis can be measured with $w = 1$ for i.i.d. symbols only.

Fig. 2 shows the 4th and 6th windowed standardized moments of various modulation formats. For i.i.d. symbols (dashed lines) in the continuous Gaussian, PCS 64-QAM, uniform 64-QAM, and uniform quadrature phase-shift keying (QPSK) systems, the windowed standardized moments are constant for all $w$, as proven in Appendix. The i.i.d. continuous Gaussian produces the greatest moments among all modulation formats, with $\bar{\mu}_4 = 2$ and $\bar{\mu}_6 = 6$. Plugging these into (15), we have that $\eta_{SPM} = \kappa_1$, implying that the EGN model degenerates to the GN model. On the other hand, the i.i.d. QPSK yields the smallest moments $\bar{\mu}_4 = \bar{\mu}_6 = 1$, thereby making the negative correction term (A) of (4) smallest and minimizing the modulation-dependent NLI.

In Fig. 2, the curves for correlated symbols are obtained with Monte-Carlo simulations, using 1D-shaping (dash-dotted lines) and 4D-shaping (solid lines) with four difference block lengths $n = 5, 10, 20, 40$. Remarkably, for these correlated symbols, the windowed central moments decrease as $w$ increases. This is due to the fact that there exist local energy structures in sphere-shaped symbols and that the proposed statistical measure can capture these local structures. For example, when we use $n = 5$ for sphere shaping that yields the square-root power distribution shown in Fig. 1(b), only small power deviations can be observed if the measurement window length $w$ is greater than $n = 5$, due to the concentration of the power near a fixed value. In Fig. 2(a), the crossing point of the i.i.d. uniform 64-QAM and sphere-shaped 64-QAM appear approximately at $w = n$. When measured with a longer window than $n$, the sphere-shaped symbols produce even a smaller kurtosis than i.i.d. symbols, which cannot be observed with the conventional measurement method. This shows that to best utilize the energy structure to reduce the windowed kurtosis, we need to put a block of sphere-shaped symbols in the shortest possible time, and this explains the reason why we map four consecutive sphere-shaped 1D symbols onto one PDM symbol. As expected, the 4D-shaped symbols produce much smaller moments than the 1D-shaped symbols at $w = 1$ (i.e., when measured per 4D symbol) due to the additional energy constraint imposed on a time slot basis. Interestingly, however, the 4D-shaped symbols can produce greater moments than 1D-shaped symbols when measured with very large window lengths, as shown in the insets in Fig. 2. This happens because, as mentioned earlier, 4D-shaping in some cases should use a greater $E^*_{\text{Shaped}}$ than 1D-shaping. This suggests that which of the 1D- and 4D-shaping is more advantageous for reducing NLI may depend on the window length, as we will discuss in the following sections.

*B. Optimal measurement window*

In [28], it has been shown that the correlation between $\bar{\mu}_4$ and NLI is maximized when $\bar{\mu}_4$ is measured with window lengths

$$w_{SPM} = 2R_{Sym}B_{Ch}|\beta_2|L_{Span}N_{Span} \qquad (16)$$

and

$$w_{XPM} = 2R_{Sym}B_{Ch}\sqrt{\Delta f \cdot N_{Ch}}|\beta_2|L_{Span}N_{Span}, \qquad (17)$$

where the window lengths are obtained separately for SPM and XPM, and where $R_{Sym}$ is the symbol rate in Baud, $B_{Ch} \approx R_{Sym}$ is the channel bandwidth in Hz, $\beta_2$ is the fiber dispersion coefficient in s²/m, $L_{Span}$ is the span length in m, $N_{Span}$ is the number of spans, and $\Delta f$ is frequency separation between adjacent channels, normalized to $R_{Sym}$ (e.g., $\Delta f = 1$ for systems with ideal Nyquist pulse shaping and zero spectral margins between channels). The window lengths were obtained in a densely packed WDM system with a total bandwidth of 100 GHz, where all channels have the same $R_{Sym}$, $B_{Ch}$, and $\Delta f$, and Erbium doped fiber amplifiers (EDFAs) are used to recover optical power after every span. It is surprising to see that the measurement window lengths given as (16) and (17) are consistent with the SPM and XPM characteristic times [3], [30], [32]–[34]. The window lengths to measure kurtosis increase linearly with the total accumulated chromatic dispersion. Under Nyquist pulse shaping, we have $B_{Ch} \approx R_{Sym}$, and thus (16) and (17) show that just as the channel memory in number of symbols due to chromatic dispersion increases quadratically with $R_{Sym}$ [30], so does the window length quadratically with $R_{Sym}$. The window length $w_{XPM}$ for XPM increases as the square root of the number of WDM channels. This square-root relation comes from the fact that distant channels contribute less to NLI [32], [33], [34, Fig. 5].

TABLE I
SIMULATION SETUP

| Name | Value |
|---|---|
| Modulation | PDM PCS 64-QAM |
| Shaping principle | 1D- or 4D-sphere shaping |
| Shaping algorithm | ESS |
| Shaping rate $H$ | 1.6 bits per positive 1D symbol |
| Shaping block length $n$ | 5, 10, 20, 40 (in 4D symbols) |
| Total bandwidth | 500 GHz |
| Center frequency | 193.4 THz (1550.1 nm) |
| Symbol rate | 5.5 or 88 GBd |
| WDM channels | 80 or 5 |
| Fiber type | SSMF |
| Fiber length | 60 km |
| Attenuation | 0.2 dB/km |
| Dispersion $\beta_2$ | 2.199×10$^{-26}$ s$^2$/m |
| Nonlinearity $\gamma$ | 1.3 ×10$^{-3}$ W$^{-1}$/m |
| Launch power | Optimized for maximum GMI |
| Amplification | EDFA |
| Noise figure | 4.5 dB |

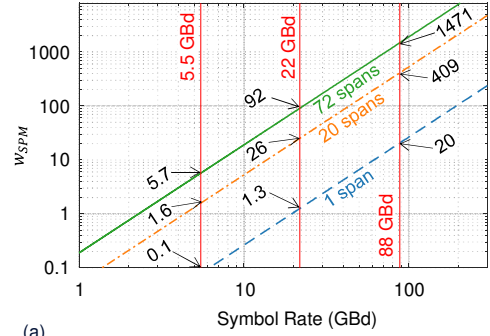

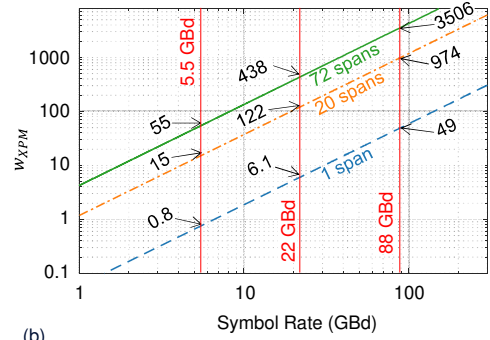

Fig. 3. Optimal window lengths to measure kurtosis for analytical estimation of NLI due to (a) SPM and (b) XPM.

The window lengths given as (16) and (17) were found in [28] by split-step simulations with a wide range of system configurations $R_{Sym} \in [1.375, 88]$ GBd, $B_{Ch} \in [1.375, 88]$ GHz, $N_{Ch} \in [1, 64]$, $\beta_2 \in [0, 2.199] \times 10^{-26}$ s$^2$/m (i.e., zero dispersion to the dispersion of SSMF), $N_{span} \in [2, 250]$ in a fixed total WDM bandwidth of 100 GHz, using PCS 16-QAM. In this work, we perform split-step simulations using the same wide range of system configurations as in [28] and the same PCS 16-QAM, except that the total bandwidth is increased to 500 GHz, accommodating 5 times more WDM channels. With this, we confirm that the window lengths given as (16) and (17) also apply to systems with different total bandwidths. As it requires a lengthy explanation, we do not provide the details on how to find the optimal window lengths for measuring kurtosis in this paper; interested readers are referred to [28], especially Fig. 3(a) and Supplementary Figs. 5-8 of [28], which were observed almost identically in this work at a total bandwidth 5 times wider than in [28].

## V. SPLIT-STEP AND EGN SIMULATIONS

We used a family of 1D-shaped PCS 16-QAM symbols in [28] to verify the dependency of NLI on the windowed moments, in which case the windowed moments are changed only by the shaping block length (cf. Fig. 2, dash-dotted lines). In this work, in addition to validating the findings of [28] with the same modulation format (PCS 16-QAM) at a wider total bandwidth as mentioned above, we also perform simulations using a larger modulation format (PCS 64-QAM) and extend the shaping method to 4D-shaping at three selected symbol rates of 5.5, 22, and 88 GBd. Since PCS 16-QAM and PCS 64-QAM, and 1D- and 4D-shaping each produce different moments at the same shaping block lengths (cf. Fig. 2, compare dashed-dotted and solid lines), this allows us to see if the windowed kurtosis leads to accurate NLI estimation for a broader class of correlated symbols.

We perform split-step simulations and compare the results with Monte-Carlo integration-based semi-analytic EGN modeling results. In a fixed total bandwidth of 500 GHz, we transmit signals at 5.5 GBd in 80 channels, at 22 GBd in 20 channels, or at 88 GBd in 5 channels. The WDM channels are uniformly spaced in 500 GHz, with the same normalized channel separation of $\Delta f = 100/88 \approx 1.136$ in all cases. Root-raised cosine (RRC) pulse shaping is used with 10% roll off. Launch power is increased from $-2$ dBm to 3 dBm per 100 GHz bandwidth in 1 dB increments, and the one that maximizes the generalized mutual information (GMI) [19] is chosen as the optimal launch power. The system parameters used for simulations are summarized in Table 1. Note that we use a nonlinear coefficient $\gamma$ of $1.3 \times 10^{-3}$ W$^{-1}$/m in Section V of this work, as compared to $1.45 \times 10^{-3}$ W$^{-1}$/m of Section IV and [28].

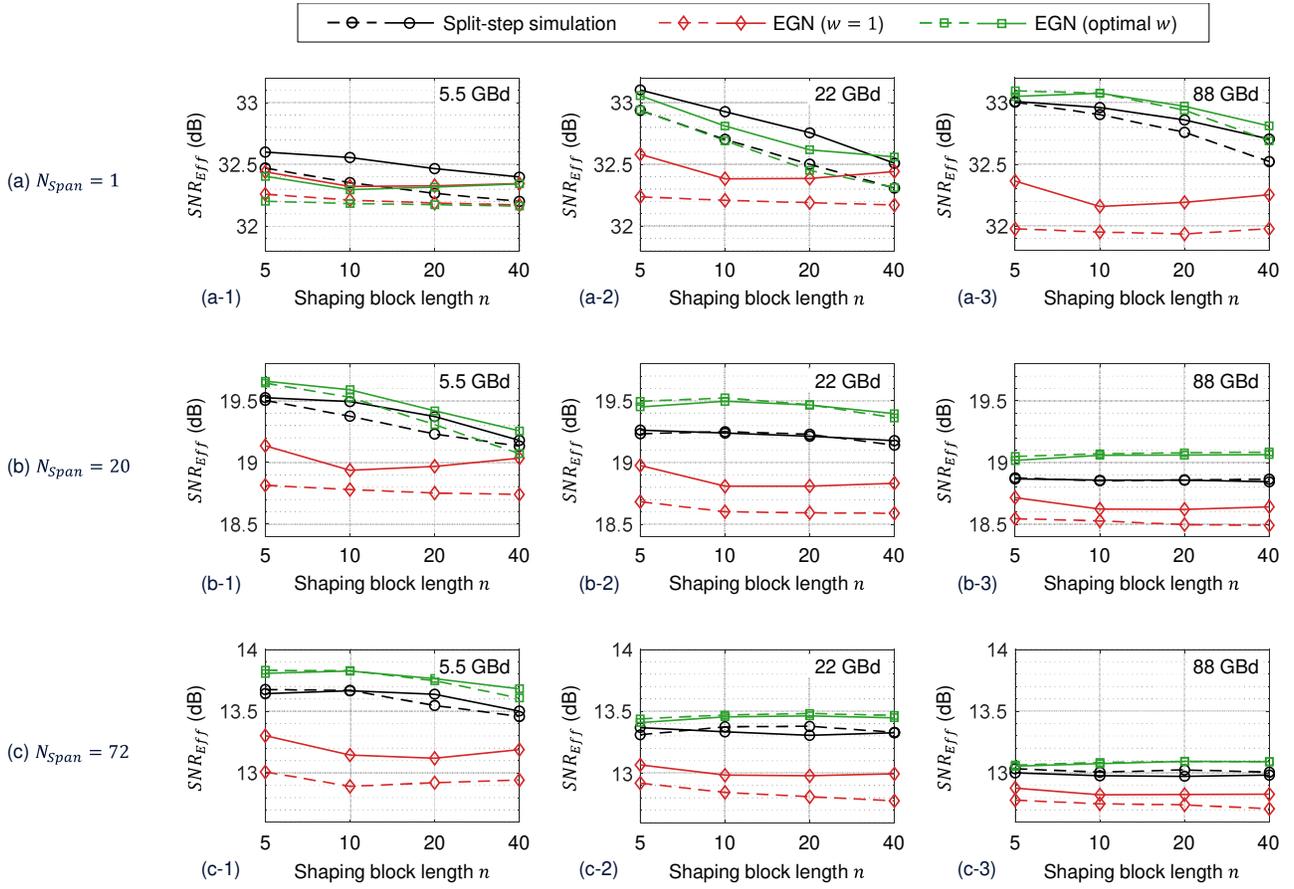

Fig. 4. Results of the split-step simulation (black circles) and the EGN model with $w = 1$ (red diamonds) and optimal $w$ (green squares) at (a) 1 span, (b) 20 spans, and (c) 72 spans. The dashed lines indicate conventional 1D-shaping, and the solid lines indicate 4D-shaping. The figures from left to right are obtained using $R_{Sym} = 5.5$, 22, and 88 GBd in order.

The optimal window lengths for three different transmission distances, at 1, 20, and 72 spans (corresponding to 60, 1200, and 4320 km), are depicted in Fig. 3 as a function of the symbol rate, obtained using (16) and (17). Note that the x- and y-axes are both logarithmic in scale. For the three symbol rates under test (red vertical lines), the exact window lengths are explicitly noted in the figure. When the accumulated dispersion is low, i.e., with a lower symbol rate of 5.5 GBd at a shorter distance of 1 span (blue dashed lines) or 20 spans (orange dash-dotted lines), the window lengths are not much greater than 1. This implies that the conventional kurtosis with $w = 1$ may produce good accuracies when predicting NLI through the EGN model. On the other hand, as the dispersion increases by either the transmission distance or the symbol rate, the window length becomes much larger than 1, and thus the conventional kurtosis and windowed kurtosis may lead to substantially different results. In Fig. 3, due to the quadratic relation with the symbol rate, the window lengths increase much faster with increasing symbol rate than with increasing distance. At the high symbol rate of 88 GBd, the window lengths reach several dozens even at 1 span, and at 20 or 72 spans, the window lengths reach hundreds to thousands. Note that if the window length exceeds several hundred, the windowed kurtosis is nearly saturated to the lowest value for all shaped symbols (cf. Fig. 2).

Using the split-step simulation and EGN model, we evaluate the effective SNR defined as

$$SNR_{Eff} = \frac{\langle \|x\|^2 \rangle}{\langle P_{ASE} \rangle + \langle P_{NLI} \rangle}, \qquad (18)$$

where the numerator quantifies the average signal power over 2 polarizations, $\langle P_{ASE} \rangle$ denotes the average amplified spontaneous emission (ASE) noise power, and the average NLI power $\langle P_{NLI} \rangle$ is given as (3). The signal, ASE noise, and NLI powers are measured over 100 GHz bandwidth at the center of the total bandwidth of 500 GHz. In case of the split-step simulation, we can obtain $SNR_{Eff}$ and $\langle P_{ASE} \rangle$ from the simulation, and then estimate $\langle P_{NLI} \rangle$ using (18). In case of the EGN model, we can obtain $\langle P_{ASE} \rangle$ and $\langle P_{NLI} \rangle$ from the model, and then estimate $SNR_{Eff}$ using (18). Fig. 4 shows the effective SNR obtained using split-step simulation (black circles), and EGN model with $w = 1$ (red diamonds) and optimal $w$ (green squares). The figures from left to right are obtained with $R_{Sym} = 5.5$, 22, and 88 GBd in order. From the figures, several observations can be made as follows:

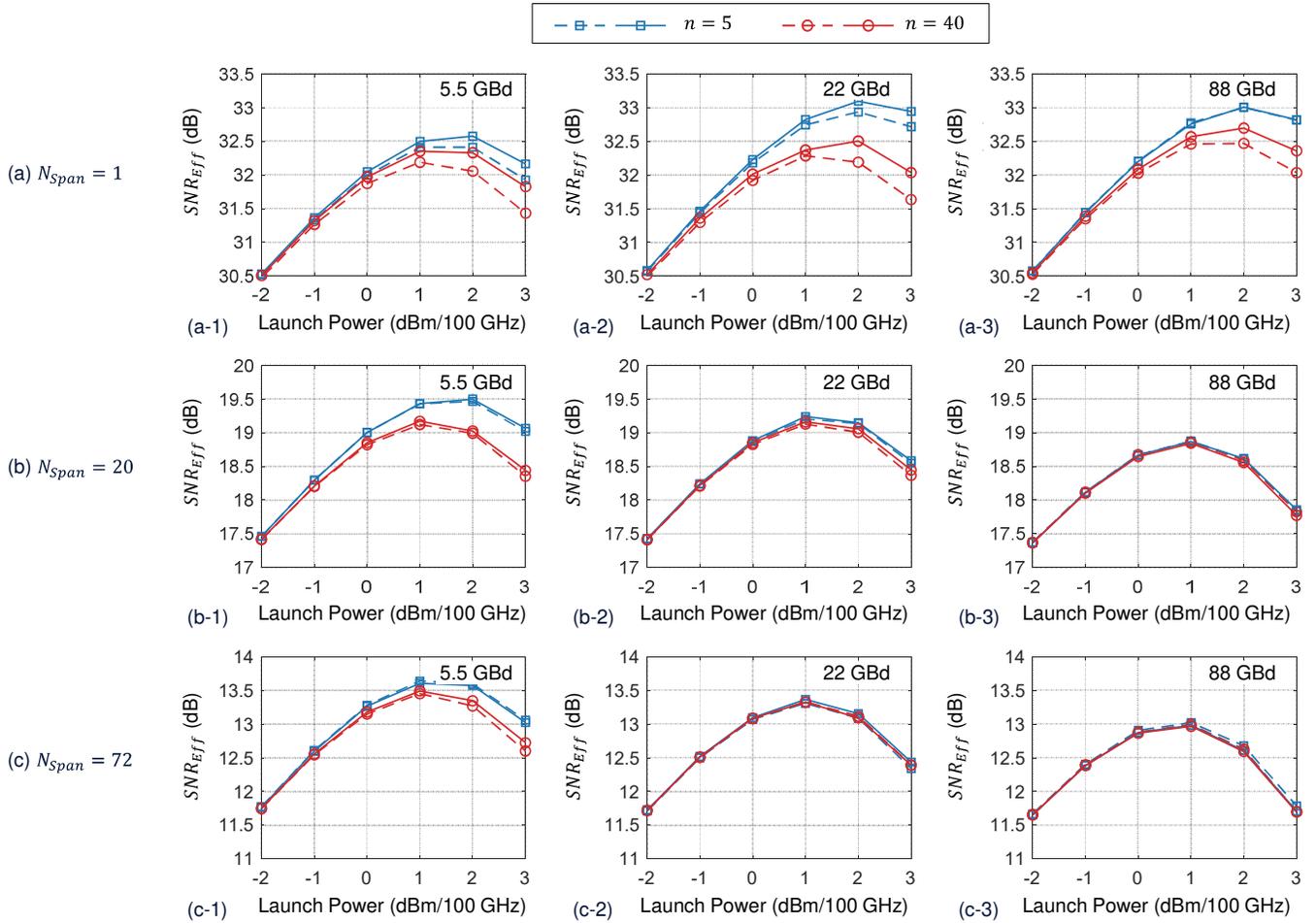

Fig. 5. Effective SNR as a function of the launch power at (1) 1 span, (b) 20 spans, and (c) 72 spans, obtained with split-step simulation. The dashed and solid lines indicate, respectively, 1D- and 4D-shaping. The figures from left to right are obtained using $R_{Sym}$ = 5.5, 22, and 88 GBd in order.

1) In general, windowed moments (green squares) produce results that are much more consistent with the split-step simulation (black circles) than conventional moments (red diamonds).
2) In Figs. 4(a-2), (a-3), (b-1), (b-2), and (c-1), the decrease in $SNR_{Eff}$ with increasing shaping block length $n$ can only be observed when the windowed moments are used, and not with the conventional moments. This change in $SNR_{Eff}$ is consistent with the fact that at the same window length determined for the underlying system configuration, the windowed kurtosis increases with the shaping block length, unless the window length is too large (cf. Fig. 2).
3) In Figs. 4(b-3), (c-2), and (c-3), despite using the windowed kurtosis, the predicted $SNR_{Eff}$ is nearly constant across all shaping block lengths. This can be explained by the fact that for the high symbol rates of 22 and 88 GBd after 20 or 72 spans, the window lengths are on the order of hundreds to thousands (cf. Fig. 3), and at these long window lengths, all the shaping block lengths under test yield nearly identical windowed kurtosis (cf. Fig. 2).
4) In the left-hand side figures of Fig. 4 where the lower symbol rate of 5.5 GBd is used, the split-step simulation results (black circles) show that 4D-shaping (solid lines) can produce higher $SNR_{Eff}$ than conventional 1D-shaping (dashed lines). The advantage of 4D-shaping over 1D-shaping is greatest when the accumulated dispersion is smallest, i.e., at 1 span. This is consistent with the fact that the 4D-shaping produces a much lower windowed kurtosis than 1D-shaping when the window length is small (cf. Fig. 2). Here, the EGN model achieves good prediction accuracies if optimal $w$ is used, but the accuracy deteriorates if the accumulated dispersion is too small to satisfy the Gaussian assumption on NLI as in Fig. 4(a-1).
5) On the other hand, in the right-hand side figures of Fig. 4, where the highest symbol rate of 88 GBd is used, the split-step simulation results (black circles) show that 4D-shaping (solid lines) produces similar or even lower $SNR_{Eff}$ than 1D-shaping (dashed lines, see also Fig. 5 for comparison between 5.5 GBd and 88 GBd at 1 and 72 spans). In this case, the EGN model gives results that are consistent with the split-step simulation when using the windowed kurtosis but provides the opposite results when using the conventional kurtosis. The reason why 4D-shaping does not perform better than 1D-shaping can be explained by the fact that the windowed kurtosis of 4D-shaped symbols becomes similar or slightly larger than that of 1D-shaped symbols, when measured with the very large window lengths for 88 GBd. On the other hand, the conventional kurtosis of 4D-shaped symbols is still much smaller than that of 1D-shaped symbols, regardless of how large the dispersion becomes for large symbol rates over long distances, and thus the EGN model incorrectly predicts with the conventional kurtosis. This shows that the conventional kurtosis minimization approach for designing modulation formats can mislead. In order to accurately evaluate the NLI tolerance of modulation formats, the windowed kurtosis with different window lengths should be used depending on various factors such as the symbol rate, transmission distance, and fiber dispersion parameter.

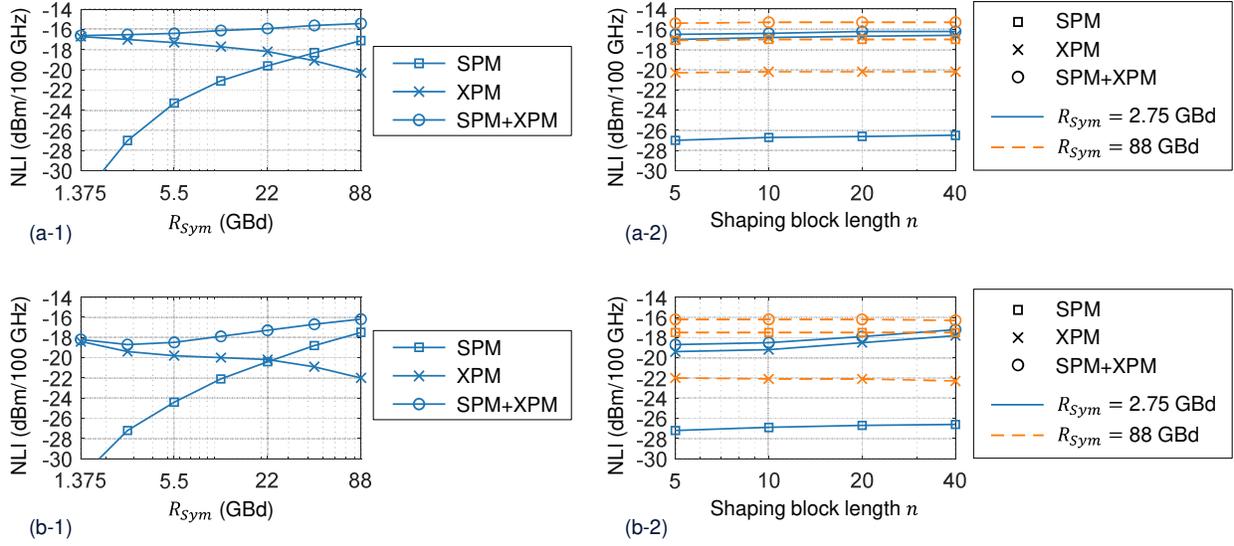

Fig. 6. NLI power obtained by EGN model with (a) $w = 1$ and (b) optimal $w$, at 72 spans with a launch power of 1 dBm/100 GHz. (a-1, b-1) NLI power as a function of $R_{Sym}$ at a fixed $n = 5$. (a-2, b-2) NLI power as a function of the shaping block length $n$ at $R_{Sym} = 2.75$ GBd (blue solid lines) and 88 GBd (orange dashed lines). Square, cross, and circle markers indicate SPM, XPM, and SPM+XPM, respectively.

Fig. 6 shows the contribution of SPM and XPM to the overall NLI, obtained by using the EGN model with $w = 1$ (Fig. 6(a)) and optimal $w$ (Fig. 6(b)), at 72 spans with a launch power of 1 dBm/100 GHz. PCS 64-QAM with 1D-shaping that has the kurtosis shown in Fig. 2 is used, and the same WDM configuration rule as above is applied for the total bandwidth of 500 GHz, with 7 different $R_{Sym} \in [1.375, 88]$ GBd. The NLI power is evaluated for the central 100 GHz bandwidth. When evaluated using the optimal $w$ at a fixed shaping block length $n = 5$ (Fig. 6(b-1)), as $R_{Sym}$ decreases, XPM (crosses) increases and SPM (squares) decreases, and the overall NLI (circles) is minimized at $R_{Sym} \approx 2.75$ GBd. However, when evaluated using $w = 1$ (Fig. 6(a-1)), the NLI is minimized at a lower symbol rate $R_{Sym} \leq 1.375$ GBd. When evaluated using the optimal $w$ with various shaping block lengths $n$ (Fig. 6(b-2)), at a high symbol rate of 88 GBd, SPM and XPM are almost constant regardless of $n$, whereas at a lower symbol rate of 2.75 GBd at which the overall NLI is approximately minimized, they both decrease as $n$ decreases. This decreasing SPM and XPM are not observed with $w = 1$ (Fig. 6(a-2)).

VI. CONCLUSION

In our previous work [28], we showed for the first time that the nonlinear fiber propagation effect of correlated symbols can be analytically determined. In this work, we demonstrated that the analytical prediction of NLI applies to more general systems than in [28], by using a much wider total system bandwidth and a different family of correlated symbols. Instead of generalizing the EGN model to account for the nonlinear fiber propagation of a multitude of correlated symbols, which appears to be a mathematically challenging task even when 4 symbols are correlated [25], [26], we allowed the model mismatch and modified the statistical measure to capture the local properties of long correlated symbols. Remarkably, the optimal window length for measuring the local properties of signal has been shown to be consistent with the current understanding of nonlinear fiber propagation phenomena.

Through split-step simulations, we demonstrated that compared to 1D-shaping, 4D-shaping designed for smaller (conventional) kurtosis does not necessarily yield better nonlinear performance in all system configurations. This cannot be explained by existing theories but can be explained by the proposed windowed statistical moments. Our findings suggest that, although it is an open question whether this is possible, the modulation format should be designed to produce a small windowed kurtosis universally for all window lengths to achieve good nonlinear tolerance in various system configurations. If this is not possible, it may be necessary to design and use a different modulation format depending on the target system configuration.

APPENDIX

Let $X, X_i$ for $i = 1, 2, \ldots$ be a sequence of i.i.d. random variables. Then, we have

$$\langle \langle X \rangle_w^k \rangle = \left\langle \left( \frac{1}{w} \sum_{i=1}^{w} X_i \right)^k \right\rangle$$
$$\stackrel{(*)}{=} \frac{1}{w^n} \left\langle \sum_{i=1}^{w} X_i^k \right\rangle$$
$$= \frac{1}{w^{n-1}} \langle X^k \rangle. \quad (A1)$$

Here, the equality $(*)$ holds for $k = 2, 3$ if and only if $\langle X \rangle = \langle X_i \rangle = 0$. Therefore, by letting $X = \wp - 1$, the equality $(*)$ holds, and (A1) can be rewritten as

$$\langle (\wp - 1)^k \rangle = \langle \langle \wp - 1 \rangle_w^k \rangle \cdot w^{k-1}$$
$$= \langle (\langle \wp \rangle_w - 1)^k \rangle \cdot w^{k-1}. \quad (A2)$$

The left-hand side of (A2) equals (6), and the right-hand side of (A2) equals (12). Therefore, if $\wp$ is i.i.d., (6) equals (12), and the windowed central moment equals the non-windowed central moment for $k = 2, 3$. Here, the choice of $X = \wp - 1$ is necessary to meet the equality condition $(*)$ of (A1), allowing us to find a statistical measure that is invariant under windowing for i.i.d. symbols. This explains the reason for using the central moments of $\wp$ in this work, instead of the standardized moments of $x$.